\documentclass[11pt,a4paper]{article}

\usepackage{epsfig}
\usepackage{amsmath}
\usepackage{amssymb}
\usepackage{verbatim}
\usepackage{bm}
\usepackage{color}

\usepackage[footnotesize]{caption}
\usepackage[subrefformat=parens,labelformat=parens]{subfig}
\usepackage{cite}
\numberwithin{equation}{section}

\newcommand{\I}{\ensuremath{\varphi}}
\newcommand{\Ib}{\ensuremath{\overline{\varphi}}}
\newcommand{\p}{\ensuremath{\phi_{+}}}
\newcommand{\m}{\ensuremath{\phi_{-}}}
\newcommand{\Pm}{\ensuremath{\phi_{{\pm}}}}
\newcommand{\qp}{\ensuremath{q_{+}}}
\newcommand{\qm}{\ensuremath{q_{-}}}
\newcommand{\qpm}{\ensuremath{q_{{\pm}}}}
\newcommand{\pb}{\ensuremath{\overline{\phi}_{+}}}
\newcommand{\mb}{\ensuremath{\overline{\phi}_{-}}}

\begin{document}

\begin{flushleft}
DESY 14-139\\
August 2014
\end{flushleft}

\vskip 1cm

\begin{center}
{\Large\bf Inflation with Fayet-Iliopoulos Terms} 

\vskip 2cm

{Clemens Wieck, Martin Wolfgang Winkler}\\[3mm]
{\it{
Deutsches Elektronen-Synchrotron DESY, 22607 Hamburg, Germany}
}
\end{center}

\vskip 1cm

\begin{abstract}
\noindent 
Two of the most attractive realizations of inflation in supergravity are based upon the presence of a constant Fayet-Iliopoulos (FI) term. In D-term hybrid inflation it is the FI term itself which sets the energy scale of inflation. Alternatively, the breaking of a $U(1)$ symmetry induced by the FI term can dynamically generate the quadratic potential of chaotic inflation. The purpose of this note is to study the possible UV embedding of these schemes in terms of the `field-dependent FI term' related to a string modulus field which is stabilized by a non-perturbative superpotential. We find that in settings where the FI term drives inflation, gauge invariance prevents a decoupling of the modulus from the inflationary dynamics. The resulting inflation models generically contain additional dynamical degrees of freedom compared to D-term hybrid inflation. However, the dynamical realization of chaotic inflation can be obtained in complete analogy to the case of a constant FI term. We present a simple string-inspired toy model of this type.
\end{abstract}

\thispagestyle{empty}

\newpage

\tableofcontents
%

\section{Introduction}

Inflation is a promising paradigm to explain the initial conditions of the universe. In particular, hybrid inflation scenarios driven by F-terms \cite{Copeland:1994vg,Dvali:1994ms} or D-terms \cite{Binetruy:1996xj,Halyo:1996pp} have been studied extensively in the literature and provide intriguing links to UV-complete theories like string theory. In D-term hybrid inflation (DHI) the vacuum energy is determined by a constant Fayet-Iliopoulos (FI) term associated with a $U(1)$ gauge symmetry. Inflation proceeds in a false vacuum state where the slope of the inflaton potential is generated by quantum corrections. When the inflaton reaches a critical value the $U(1)$ symmetry is spontaneously broken and inflation ends in a waterfall phase transition.

On the other hand, implementations of chaotic inflation \cite{Linde:1983gd} in supergravity have gained new traction in the literature since the possible discovery of primordial gravitational waves by the BICEP2 experiment \cite{Ade:2014xna}. While the signal is currently being analyzed regarding a possible foreground contamination (see, for example, \cite{Flauger:2014qra}), the tensor-to-scalar ratio  inferred by the BICEP2 collaboration is in good agreement with the value predicted by chaotic inflation with a quadratic potential.

Recently it was noted that DHI may contain a regime of chaotic inflation \cite{Buchmuller:2014rfa}. Specifically, if the critical value of the inflaton is super-Planckian a phase of chaotic inflation may follow after the $U(1)$ phase transition. We show explicitly that DHI in the waterfall regime is identical to the standard realization of chaotic inflation in supergravity discussed in \cite{Kawasaki:2000yn,Kallosh:2010ug}. The role of the ``stabilizer field'' invoked in \cite{Kawasaki:2000yn,Kallosh:2010ug} is played by one of the waterfall fields. The non-minimal K\"ahler potential of the stabilizer, which is required to decouple it from inflation, is explained by a $U(1)$ gauge interaction and can be obtained by integrating out the vector supermultiplet of the broken symmetry.

With regard to a possible UV embedding of these inflation models, it was noted in \cite{Komargodski:2009pc,Komargodski:2010rb,Dienes:2009td} that constant FI terms in supergravity are potentially troubled. However, supergravity models in which the arguments of \cite{Komargodski:2009pc,Komargodski:2010rb,Dienes:2009td} do not apply have been studied in \cite{Seiberg:2010qd,Distler:2010zg,Catino:2011mu}. Given this ongoing discussion in the literature, we are particularly interested in `field-dependent FI terms'\footnote{Notice that this terminology is somewhat misleading. The `field-dependent FI term' is the D-term of a modulus field which transforms non-linearly under a $U(1)$ symmetry. Thus, it is quite different in nature from the constant gauge-invariant term introduced by Fayet and Iliopoulos.} generated in the presence of a (pseudo-)anomalous $U(1)$ symmetry, in the following denoted by $U(1)_\text{A}$. The appearance of such D-terms was first discussed in the context of the Green-Schwarz mechanism \cite{Green:1984sg} in heterotic string theory in \cite{Dine:1987xk}. There it was argued that the dilaton, whose axionic part cancels the anomalies associated with $U(1)_\text{A}$, acquires a D-term which bears resemblance to an FI term if the dilaton is assumed to be stabilized. Similar D-terms arise in certain compactifications of type IIB string theory, where the role of the dilaton is played by a K\"ahler modulus (see, for example, the discussion in \cite{Achucarro:2006zf}). However, it was soon realized that modulus or dilaton stabilization is a subtle issue in the presence of the field-dependent FI term \cite{Binetruy:2004hh}. Gauge invariance of the modulus superpotential poses severe restrictions on possible setups \cite{Dudas:2005vv,Choi:2005ge,Villadoro:2005yq,Achucarro:2006zf}. In particular, it has been shown that invoking non-perturbative superpotential terms for the K\"ahler modulus or dilaton requires the inclusion of additional fields charged under $U(1)_\text{A}$. Otherwise, the respective field can not be stabilized in a gauge-invariant way. This can be achieved, for example, by including a gauge sector with chiral matter which undergoes gaugino condensation \cite{Taylor:1982bp,Affleck:1984xz}.

We wish to clarify whether a field-dependent FI term can play the role of an effective constant which drives inflation.\footnote{For alternative and recent attempts to reconcile inflation with field-dependent FI terms, see \cite{Gwyn:2011tf,Hebecker:2012aw,Li:2014owa,Li:2014xna,Li:2014lpa}.} In the case of DHI we discuss a series of obstacles which prevent possible setups from resembling the simple controllable model introduced in \cite{Binetruy:1996xj,Halyo:1996pp}. Taking stabilization of all additional fields into account, it turns out that in all feasible setups of modulus stabilization with non-perturbative superpotentials the modulus never decouples from the dynamics of inflation, rendering much more complicated multi-field inflation models. We remark, however, that in cases where the modulus which generates the FI term does not appear in the superpotential some of our arguments may be avoided. This can be realized, for example, in the Large Volume Scenario, cf.~the discussion in \cite{Hebecker:2012aw}.

In the case of inflation in the chaotic regime of DHI, on the other hand, a separation of the modulus from the inflaton dynamics seems possible in all moduli stabilization schemes. We provide an example in which the effective theory, after integrating out the modulus and the heavy $U(1)$ vector supermultiplet supersymmetrically, is identical to single-field chaotic inflation.

%

\section{FI Terms in Supergravity and String Theory}
\label{sec:FIterms}

In order to introduce the basic notions for the following discussion, let us briefly review both constant FI terms and field-dependent FI terms related to an anomalous $U(1)$ symmetry. 

%
\subsection{Constant FI Terms}
\label{sec:ConstantFI}

In a supergravity theory with $U(1)$ gauge interactions the Lagrangian is determined by the choice of superpotential $W$, K\"ahler potential $K$, gauge kinetic function $f$, and Killing vectors $\eta^\alpha$ specifying the gauge transformation properties of chiral superfields $\phi_\alpha$. The superpotential and K\"ahler potential enter the Lagrangian in the combination $K + \log |W|^2$, which must be gauge-invariant. The gauge kinetic function transforms trivially under the $U(1)$ up to a possible shift required for anomaly cancellation. It determines the gauge coupling as ${g^2 = (\text{Re}\, f)^{-1}}$. In case the $U(1)$ symmetry is linearly realized, chiral superfields $\phi_\alpha$ transform as
\begin{align}\label{eq:ChiralTransform}
\phi_\alpha \to e^{i q_\alpha \epsilon } \phi_\alpha\,,
\end{align} 
where $\epsilon$ is a chiral superfield gauge transformation parameter and $q_\alpha$ denotes the charge of $\phi_\alpha$. This corresponds to the choice of Killing vector $\eta^\alpha = i q_\alpha \phi_\alpha$. The transformation of the $U(1)$ vector superfield $\mathcal V$ can be written as
\begin{align}
\mathcal V \to \mathcal V - \frac{i}{2} (\epsilon -\bar \epsilon)\,.
\end{align}
The scalar potential may contain an F-term and a D-term piece, i.e., $V = V_F + V_D$ with
\begin{align}
V_F &= e^K \left(K^{\alpha \bar \alpha} D_{\alpha} W D_{\bar \alpha} \overline W - 3 |W|^2 \right) \,, \\
V_D &= \frac{1}{2 \text{Re}\, f} D^2\,. \label{eq:DPot}
\end{align}
The $D$-terms associated with the $U(1)$ can be expressed as
\begin{align}
D = - i \eta^\alpha K_\alpha  \underbrace{- i\frac{W_\alpha}{W} \eta^\alpha}_{\equiv\,\xi}\,.
\end{align}
Notice that, by gauge invariance, $W$ may transform with a constant phase denoted by $\xi$. This is precisely the constant FI term introduced in \cite{Fayet:1974jb}.

%
\subsection{Field-Dependent FI Terms}
\label{sec:FielddepFI}

The consistency of constant FI terms from the perspective of string theory and quantum gravity is an issue of vital discussion in the literature. In \cite{Komargodski:2009pc,Dienes:2009td,Komargodski:2010rb} it was pointed out that a constant FI term in supergravity may be inconsistent when coupled to quantum gravity, while possible counter-examples have been studied in \cite{Seiberg:2010qd,Distler:2010zg,Catino:2011mu}. Whatever the outcome of this discussion, so far there are no known four-dimensional effective theories derived from string theory which contain constant FI terms. String theory, however, provides an elegant mechanism which generates field-dependent FI terms which, from the viewpoint of cosmology, may play a similar role as their constant counterparts.

Depending on the full gauge group and chiral spectrum of the theory under consideration, a $U(1)$ symmetry like the one in Section~\ref{sec:ConstantFI} can have several gauge anomalies, in which case we denote it by $U(1)_\text{A}$. These manifest as divergences of the gauge current $J$, i.e.,
\begin{align}
\partial_\mu J^\mu \propto c_1 A_{G^2-U(1)_\text{A}} \,\text{tr}\, \mathcal{F}_{\mu \nu} \tilde{ \mathcal F}^{\mu \nu} + c_2 A_{U(1)_\text{A}^3} \, F_{\mu \nu} \tilde F^{\mu \nu} + c_3 A_{\text{grav}^2-U(1)_\text{A}}\,  \text{tr}\, R_{\mu \nu} \tilde R^{\mu \nu}\,,
\end{align}
where $\mathcal F$, $F$, and $R$ denote the field strengths of a non-Abelian gauge group piece $G$, $U(1)_\text{A}$, and the Riemann tensor, respectively. The prefactors $c_i$ depend on the underlying string construction, while the anomaly coefficients $A$ are given by
\begin{align}\label{eq:anomcoeffs}
A_{G^2-U(1)_\text{A}} = \sum_f q_f \ell ({\textbf R}_f) \,, \quad A_{U(1)_\text{A}^3} = \sum_\alpha q_\alpha^3\,,
\quad  A_{\text{grav}^2-U(1)_\text{A}}= \sum_\alpha q_\alpha \,. 
\end{align}
The first sum runs over all chiral fermions transforming in the representation $\textbf{R}$ of $G$ and $\ell (\textbf R)$ denotes the quadratic index of $\textbf R$. The sums in the second and third expression run over all chiral fermions.

For the theory to be consistent, all anomalies must be canceled by the four-dimensional variant of the Green-Schwarz mechanism \cite{Green:1984sg}. This means there must be at least one axion which shifts under $U(1)_\text{A}$, and this shift cancels all anomalies via its coupling to the field strengths. Motivated by compactifications of type IIB string theory, we take the axion to be the imaginary part of a K\"ahler modulus $\rho$ and assume all other moduli to be stabilized by fluxes \cite{Giddings:2001yu}. Note that the discussion proceeds analogously in heterotic string theory with the dilaton playing the role of the K\"ahler modulus. The transformation of $\rho$ under $U(1)_\text{A}$ reads
\begin{align}\label{eq:Ttransform}
\rho \to \rho - i \delta_\text{GS} \epsilon\,,
\end{align}
which corresponds to the Killing vector $\eta^\rho = - i \delta_\text{GS}$. In what follows we consider the case $G = SU(N_\text{c})$ and $N_\text{f}$ quark pairs transforming as $(N_\text{c},q)$ and $(\bar N_\text{c},\tilde q)$ under $SU(N_\text{c}) \times U(1)_\text{A}$, respectively. Cancellation of the pure $U(1)_\text{A}^3$ and the mixed $SU(N_\text{c}) \times U(1)_\text{A}^2$ anomaly then implies \cite{Achucarro:2006zf}
\begin{align}\label{eq:DeltaGS}
\delta_\text{GS} = \frac{1}{6 \pi \kappa} \sum_\alpha q_\alpha^3 = \frac{1}{4 \pi \tilde \kappa} N_\text{f} (q + \tilde q)\,,
\end{align}
where the first sum again runs over all chiral fermions. We do not impose additional constraints on $\delta_\text{GS}$ related to the cancellation of the gauge-gravity anomaly, as in type IIB orientifold compactifications the coupling of the axion to the Riemann tensor is model-dependent. The coefficients $\kappa$ and $\tilde \kappa$ which enter the above equation are $\mathcal{O}(1)$ constants which appear in the gauge kinetic functions, i.e.,
\begin{align}
f = \frac{\kappa}{2 \pi} \rho\,, \qquad \tilde f = \frac{\tilde \kappa}{2 \pi} \rho\,, 
\end{align}
for $U(1)_\text{A}$ and $SU(N_\text{c})$, respectively. The $U(1)_\text{A}$ gauge coupling is given by 
\begin{align}
g^2 = \frac{1}{\text{Re}\, f} = \frac{4 \pi}{\kappa (\rho + \bar \rho)}\,,
\end{align}
and similarly for the gauge coupling of the $SU(N_\text{c})$. In the following we choose a normalization which coincides with the one used in the work of KKLT \cite{Kachru:2003aw}, i.e., $\kappa = \tilde \kappa = \frac12$.

Since $\rho$ transforms non-trivially under $U(1)_\text{A}$, the familiar no-scale K\"ahler potential must be modified accordingly, 
\begin{align}
K = -3 \log{\left(\rho + \bar \rho \right)} \quad \longrightarrow  \quad K = -3 \log{ \left(\rho + \bar \rho - 2 \delta_\text{GS} \mathcal V \right)}\,.
\end{align}
Allowing for the presence of additional chiral fields $\phi_\alpha$ which transform linearly under $U(1)_\text{A}$, the D-term potential reads
\begin{align}\label{eq:DPotSpec}
V_D = \frac{4 \pi}{\rho + \bar \rho} \left(  \sum_\alpha q_\alpha K_\alpha \phi_\alpha + \xi_\text{GS}\right)^2\,,
\end{align}
where we have assumed gauge invariance of $W$, i.e., the absence of a constant FI term in $V_D$. The piece
\begin{align}\label{eq:xiGS}
\xi_\text{GS} \equiv - \delta_\text{GS}\, \partial_{\rho} K \simeq \frac{3 \delta_\text{GS}}{\rho + \bar{\rho} }\,,
\end{align}
is usually called a field-dependent FI term in the literature.

%

\section{Inflation with Constant FI Terms}
\label{sec:Inflation}

Before discussing inflation in models with a field-dependent FI term as in Eq.~\eqref{eq:DPotSpec}, let us first turn to the simpler case of inflation with constant FI terms. The prime example of this kind is D-term hybrid inflation. We first review the well-known embedding of DHI in supergravity before discussing a very interesting and less investigated situation: DHI can contain a phase of chaotic inflation with a quadratic potential after the waterfall transition.

%
\subsection{D-Term Hybrid Inflation}
\label{sec:DHI}

DHI in supergravity can be described by the superpotential
\begin{align}
W = \lambda\I\p\m \,,
\end{align}
and K\"ahler potential
\begin{align}
K = |\p|^2 + |\m|^2 - \frac{(\varphi-\bar{\varphi})^2}{2} \,.
\end{align}
Here, the real part of $\varphi$ is identified with the inflaton, protected from supergravity corrections by a shift symmetry of $K$, and $\Pm$ are the waterfall fields responsible for ending inflation. They carry the charges $\qpm$ under a $U(1)$ gauge symmetry. Along the inflationary trajectory $\I=\Ib$, the F- and D-term potentials read 
\begin{align}\label{eq:VFVD}
V_F &= \lambda^2\,e^{|\phi_-|^2}\, |\phi_-|^2 \varphi^2\,,\nonumber\\
V_D &= \frac{g^2}{2} \left(q_- |\phi_-|^2 + \xi \right)^2\,,
\end{align}
where we have set $\p=0$ which corresponds to its minimum during and after inflation. Notice that gauge invariance requires $\qp+\qm=\xi$. For the moment, we have neglected the dependence of the K\"ahler potential on the vector superfield of the $U(1)$.

The scalar potential $V = V_F + V_D$ has a supersymmetric Minkowski minimum at \linebreak $|\m|=\sqrt{\xi/|\qm|}$ and $\I=0$. For large values of the inflaton field, $\I > \I_\text{c} \equiv  g \sqrt{|\qm|\xi} /\lambda$, the potential has a plateau where $\m=0$, and the gauge symmetry is restored. The corresponding potential energy is determined by the FI term,
\begin{align}
V_0 = \frac{g^2 \xi^2}{2}\,.
\end{align}
The Yukawa interaction in the superpotential breaks the shift symmetry in the inflaton direction and lifts the potential at the one-loop level, generating a slope for the inflaton. Standard DHI has a potential problem due to the generation of cosmic strings in the $U(1)$ phase transition. Furthermore, it predicts a scalar spectral index of $n_\text{s}>0.98$ in tension with CMB data~\cite{Ade:2013uln}. However, minor modifications of the K\"ahler potential can reconcile the model with observation (see for example \cite{Buchmuller:2012ex}).

%
\subsection{Chaotic Inflation}
\label{sec:ChaoticInf}

It was realized recently that in DHI inflation not necessarily terminates after the $U(1)$ phase transition \cite{Buchmuller:2014rfa}. If the critical field value is very large, $\I_\text c\gg 1$, the scalar potential in the waterfall regime may be sufficiently flat for inflation to continue. This type of situation is achieved if the Yukawa coupling $\lambda$ is suppressed compared to the gauge coupling~$g$. Indeed, the inflaton potential~\eqref{eq:VFVD} is of the form $m^2\I^2$ close to the supersymmetric minimum which suggests the possibility of chaotic inflation.

In order to see that, in the waterfall regime, DHI is indeed identical to the standard realization of chaotic inflation in supergravity, we consider the full K\"ahler potential including the $U(1)$ vector superfield $\mathcal V$,
\begin{align}
K = \pb e^{2\qp \mathcal V} \p+\mb e^{2 \qm\mathcal V} \m - \frac{(\varphi-\bar{\varphi})^2}{2} + 2\xi \mathcal V\,.
\end{align}
We perform the field redefinitions
\begin{align}\label{eq:FieldRedefs}
 \p \rightarrow \left(\frac{\m}{\sqrt{\xi/|\qm|}}\right)^{-\frac{\qp}{|\qm|}}\, \p\,,\qquad \mathcal V\rightarrow \mathcal V+\frac{1}{2|\qm|}\log\left(\frac{|\m|^2}{\xi/|\qm|}\right)\,,
\end{align}
which, after a K\"ahler transformation, yield
\begin{align}
W &= \lambda\sqrt{\frac{\xi}{|\qm|}}\,\I\,\p \,,\\
  K &= \pb e^{2\qp \mathcal V} \p     + \frac{\xi}{|\qm|} e^{2\qm \mathcal V}  - \frac{(\varphi-\bar{\varphi})^2}{2} + 2\xi\,\mathcal V\,.
\end{align}
Apparently, the field redefinitions in \eqref{eq:FieldRedefs} correspond to a gauge choice. What we end up with is the superfield version of unitary gauge, as can be seen from the fact that the chiral superfield $\m$ has disappeared from the spectrum. It has been eaten by the vector superfield $\mathcal V$, which became massive in turn. Integrating out $\mathcal V$ supersymmetrically by solving its equation of motion, $\partial_{\mathcal V} K = 0$, yields
\begin{equation}
 \mathcal V=-\frac{\qp}{2|\qm|\xi}\left|\p\right|^2 + \mathcal{O}\left(\left|\p\right|^4\right)\,.
\end{equation}
After performing another K\"ahler transformation we arrive at the effective super- and K\"ahler potential,
\begin{align}\label{eq:effChaotW}
W &= m \I\p\\
K &= |\p|^2 - \frac{|\p|^4}{\Lambda^2} -\frac{(\varphi-\bar{\varphi})^2}{2}\,, \label{eq:effChaotK}
\end{align}
with $m=\lambda \sqrt{\xi/|\qm|}\,e^{\xi/2|\qm|}$ and $\Lambda^2=2|\qm|\xi/\qp^2$. 

Notice that Eqs.~\eqref{eq:effChaotW} and \eqref{eq:effChaotK} define the standard embedding of chaotic inflation in supergravity \cite{Kawasaki:2000yn,Kallosh:2010ug}. Here, $\p$ plays the role of the stabilizer field. Note, however, that in~\cite{Kallosh:2010ug} the K\"ahler potential term $|\p|^4/\Lambda^2$ was introduced by hand in order to give a sufficiently large mass, $m_{\p}>H$, to the stabilizer field during inflation. In our case this term arises in the effective theory via the exchange of the heavy $U(1)$ gauge boson.

So far we have worked in the supersymmetric limit, which is valid only in the vicinity of the minimum $\I=0$. However, corrections are suppressed as long as the scale of the $U(1)$ breaking is large compared to the supersymmetry breaking scale, which coincides with the Hubble scale. This becomes evident by considering the full potential. Tracing the minimum of $\m$, the inflaton potential for $\I < \varphi_\text{c}$ takes the form\footnote{We neglect the small correction to $V$ from the factor $e^K$.}
\begin{align}
V = V_0 \left( 1 - \frac{V_0}{2 g^2 \xi^2} \right)\,,
\end{align}
with $V_0 =  \lambda^2 \I^2 \xi$. This implies that, if $\sqrt{g \xi} > M_\text{GUT} \sim 0.01$ in Planck units, the last 50-60 e-folds of inflation can occur within the quadratic regime of the potential. The cosmic string problem of DHI is absent in this case since the $U(1)$ symmetry is already broken during inflation.

%

\section{Field-Dependent FI Terms and Modulus Stabilization}
\label{sec:ModStab}

The aim of the present note is to investigate whether the simple inflation models discussed in Section~\ref{sec:Inflation} can effectively arise in a UV-complete theory like string theory. Therefore, we concentrate on the field-dependent FI terms related to the Green-Schwarz mechanism in the presence of an anomalous $U(1)_\text{A}$ symmetry, as introduced in Section~\ref{sec:FielddepFI}. The field-dependent FI-term from a modulus $\rho$ scales as $(\rho +\bar{\rho})^{-1}$ and the corresponding Lagrangian scales as $(\rho +\bar{\rho})^{-3}$, cf.~Eq.~\eqref{eq:DPotSpec}. Thus, if the super- and K\"ahler potential exhibit no further dependence on $\rho$ it is a runaway direction. Therefore, an appropriate mechanism to stabilize $\rho$ has to be considered. 

Naively, we could assume that $\rho$ obtains a large supersymmetric mass $m_\rho \gg \xi_{\text{GS}}$ by some unspecified mechanism so that the field-dependent FI term becomes an effective constant. However, it was argued in~\cite{Binetruy:2004hh} that this assumption is inconsistent. The reason is that the vector superfield $\mathcal V$ of $U(1)_\text{A}$ would receive the same large mass $m_\rho$ via the St\"uckelberg mechanism. This, however, immediately implies that one can integrate out $\mathcal V$ supersymmetrically at the scale $m_\rho$ which excludes the very existence of an FI term in the effective theory. Hence, a more careful treatment of modulus stabilization is required in the presence of the field-dependent FI term.

The standard procedure to stabilize the lightest K\"ahler modulus $\rho$ is to employ instantonic contributions to the superpotential of the form
\begin{align}
W = W_0 +\sum_j A_j e^{-a_j \rho}\,.
\end{align}
The interplay of one or several such terms with a constant $W_0$, stemming from fluxes in the internal manifold \cite{Gukov:1999ya}, or with corrections to the K\"ahler potential can lead to stable minima for $\rho$.

The coefficients $A_j$ are typically assumed to be constant in the effective theory and may arise from integrating out heavy moduli. However, if $\rho$ contains the Green-Schwarz axion, constant coefficients $A_j$ would result in a violation of $U(1)_\text{A}$ gauge invariance. In order to remedy the theory, the $A_j$ must be promoted to functions $A_j(\phi_\alpha)$ of chiral superfields $\phi_\alpha$ which carry charge under $U(1)_\text{A}$. Writing each piece of the superpotential in the form
\begin{equation}\label{eq:gauginvw}
 W \supset A(\phi_\alpha) \;e^{-q_0 \rho / \delta_\text{GS}}\,,
\end{equation}
gauge invariance implies $q\left[A(\phi_\alpha)\right]=-q_0$ for the charge of the function $A$, cf.~the transformation \eqref{eq:Ttransform}. Superpotential terms as in Eq.~\eqref{eq:gauginvw} arise, for example, in intersecting D-brane models where the couplings between matter fields are suppressed by the world-sheet instanton action. Generation of Yukawa couplings of this type has first been treated in \cite{Blumenhagen:2006xt,Ibanez:2006da}, for a review see \cite{Blumenhagen:2009qh}. Alternatively, the $\phi_\alpha$ can be associated with the mesonic states of a strongly coupled non-Abelian gauge theory. Consider an $SU(N_\text{c})$ gauge theory with one pair of quarks $\{Q, \tilde{Q}\}$ transforming as $(N_\text{c},q)$ and $(\bar{N}_\text{c},\tilde{q})$ under $SU(N_\text{c})\times U(1)_\text{A}$, respectively. To ensure that the D-term potentials of these fields do not cancel the modulus-dependent FI term we assume $q+\tilde{q}>0$. The $SU(N_\text{c})$ undergoes gaugino condensation at a scale
\begin{align}
\Lambda=e^{-2 \pi \rho/(3 N_\text{c}-1)}\,.
\end{align}
At energy scales below $\Lambda$ the effective theory can be described by the (canonically normalized) mesonic degrees of freedom $M=\sqrt{2 Q \bar{Q}}$. The gauge-invariant superpotential of ~\cite{Taylor:1982bp,Affleck:1984xz} reads,
\begin{align}\label{eq:WADS}
W = (N_\text{c} - 1) \left( \frac{2\,e^{-2 (q+\tilde{q}) \rho /\delta_\text{GS}}}{M^2}  \right)^{\frac{1}{N_\text{c}-1}}\,,
\end{align}
after inserting the expression for $\delta_\text{GS}$ in Eq.~\eqref{eq:DeltaGS}. In the case of gaugino condensation the function $A$ in Eq.~\eqref{eq:gauginvw} is generically non-analytic. This is important since any field with negative $U(1)_\text{A}$ charge\footnote{Notice that exchanging `negative charge' with `positive charge' is merely a choice of convention. Only the sign relative to $\xi_\text{GS}$ is of importance.} entering $A$ can potentially cancel the FI term through its vacuum expectation value. Only for non-analytic $A$ the inclusion of negatively charged fields is unnecessary.\footnote{This fact was used in \cite{Achucarro:2006zf} to construct consistent string models with KKLT stabilization and D-term uplift.}

From the perspective of modulus stabilization the dependence of $A$ on other chiral fields is undesirable: the non-perturbative superpotential of Eq.~\eqref{eq:gauginvw} now induces couplings of the modulus to other light degrees of freedom, rather than generating a mass term. Only if the fields $\phi_\alpha$ themselves are stabilized appropriately an effective modulus mass term may arise.

%

\section{D-Term Inflation from Field-Dependent FI Terms}
\label{sec:DtermFI}

Having discussed modulus stabilization, let us analyze whether DHI can proceed with a field-dependent FI term. We are interested in situations where the modulus $\rho$ is stabilized during inflation and does not perturb the dynamics of DHI.\footnote{The back-reaction of stabilized K\"ahler moduli on DHI, in setups where $\rho$ is a gauge singlet, has been previously studied in \cite{Brax:2006yq,Buchmuller:2013uta}.} As a starting point, assuming that the superpotential explicitly depends on the charged modulus, we consider 
\begin{align}
W=  \lambda \varphi \p \m + W_\text{mod}(\rho)\,,
\end{align}
which entails the superpotential of hybrid inflation and the piece ${W_\text{mod} = A(\phi_\alpha)e^{-q_0 \rho/\delta_\text{GS}}+\dots}$ responsible for modulus stabilization. In order to promote the instanton contribution to a mass term, stabilization of the fields $\phi_\alpha$ must be achieved by one of the following mechanisms.

%
\paragraph{Vector-like mass terms}
$\,$ \\
The presence of gauge anomalies implies charged chiral states in the spectrum. However, as the $\phi_\alpha$ which enter $A(\phi_\alpha)$ constitute only a subset of the spectrum, they may not contribute to the anomaly, i.e., they could still receive large vector-like masses of the form $m \phi_\alpha \bar{\phi}_\alpha$. In this case, the $\phi_\alpha$ and the $\bar{\phi}_\alpha$ can be integrated out supersymmetrically yielding $A(\phi_\alpha)=0$. This would imply that the instanton term disappears in the effective theory below the scale $m$.\footnote{This does not hold for non-analytic functions $A$ which arise via gaugino condensation. However, in the case of gaugino condensation vector-like mass terms do not appear because the effective degrees of freedom, the mesons, are already two-particle states.}

%
\paragraph{Soft mass terms}
$\,$ \\
Soft masses for the $\phi_\alpha$ may be generated by non-vanishing F- and D-terms. If the field-dependent FI term is not canceled, gauge-mediated soft masses of the form
\begin{align}\label{eq:DSoftMass}
\mathcal L_\text{soft}^D = g^2 q_\alpha\,\xi_{\text{GS}} |\phi_\alpha|^2\,,
\end{align}
arise. In addition, depending on the mechanism of modulus stabilization and supersymmetry breaking, gravity-mediated soft terms may appear.\footnote{Notice that the inflaton, protected by a shift symmetry of the K\"ahler potential, does not receive a soft mass term at tree-level.} For a minimal choice of the K\"ahler potential, these are expected to be of the form
\begin{align}\label{eq:FSoftMass}
\mathcal L_\text{soft}^F = m_{3/2}^2 |\phi_\alpha|^2\,,
\end{align}
where $m_{3/2}=e^{K/2}W$ denotes the gravitino mass.

%
\paragraph{Mass terms from spontaneous symmetry breaking}
$\,$ \\
Finally, if $U(1)_\text{A}$ is broken spontaneously, Yukawa couplings can become effective mass terms. Consider, for example, a mesonic state $M$ with a coupling $\lambda \m M^2$ to the waterfall field. In the true vacuum of the theory $\m$ cancels the FI term and the meson receives the mass $\lambda \langle \m \rangle$.\medskip

From this discussion it is clear that modulus stabilization either requires the spontaneous breaking of $U(1)_\text{A}$ or the breaking of supersymmetry. In principle, all ingredients exist within the simple DHI setup of Section~\ref{sec:DHI}: during inflation supersymmetry is broken by the inflaton sector while the $U(1)$ symmetry is intact. After inflation supersymmetry is restored but the $U(1)$ is spontaneously broken by the vev of $\m$. While this may lead to successful stabilization of all fields, the responsible mechanism is clearly different during and after inflation. Therefore, the modulus sector does not decouple from the inflaton dynamics and we are left with an inflation model with several dynamical degrees of freedom. This may happen, for example, in the supersymmetric racetrack scheme studied in \cite{Kallosh:2004yh}.

To obtain the simple controllable DHI setup, the same mechanism of modulus stabilization must operate in the entire cosmological history. This requires the inclusion of additional sources of supersymmetry breaking which fix the modulus during and after inflation. A similar conclusion has previously been drawn in~\cite{Binetruy:2004hh}. There it was noted that, in a field-dependent realization of DHI, F-terms and D-terms must split their roles in a way that F-terms provide modulus stabilization while D-terms drive inflation. Assuming that the modulus mass is comparable to the gravitino mass, $m_\rho  \sim m_{3/2}$, as for example in the setup by KKLT \cite{Kachru:2003aw}, results in the constraint
\begin{align}\label{eq:fdterms}
 m_{3/2} >  g \xi_\text{GS}\,,
\end{align}
which ensures that the modulus decouples from inflation. In the following, we wish to point out that a series of problems arises even if this constraint is satisfied. 

First, no negatively charged fields beyond $\m$ should be introduced as these fields would receive large tachyonic masses during inflation and tend to cancel the FI term. Therefore, we consider the case of gaugino condensation, where the function $A$ in the instanton term contains only the positively charged mesonic fields, cf.~the discussion in Section~\ref{sec:ModStab}. It turns out that in this setup, condition~\eqref{eq:fdterms} is insufficient to decouple the modulus sector from inflation. This is because during inflation the FI term induces soft masses $m_{M} \sim g\sqrt{\xi_\text{GS}}$ for the meson fields.\footnote{Without loss of generality we assume $q_M \sim \mathcal O(1)$ for the $U(1)_\text{A}$ charge of the mesons.} These soft masses are enhanced compared to the Hubble scale as they originate from gauge mediation. In order to avoid that the mesons, and as a consequence also the modulus, are shifted by large amounts at the end of inflation, the gauge-mediated masses should be subdominant. This can be achieved by introducing even larger gravity-mediated soft masses $m_M\sim m_{3/2}  > g\sqrt{\xi_{\text{GS}}}$. At the same time the waterfall fields must be protected against such large gravity-mediated masses by a specific choice of K\"ahler potential, otherwise inflation would never end. The origin of this sequestering could lie in a higher-dimensional theory where the dominant source of supersymmetry breaking is localized on a different brane than the waterfall fields~\cite{Randall:1998uk}.

Second, even in this case, another type of problem occurs related to the size of the instanton term. Given that modulus stabilization must proceed via supersymmetry breaking, one expects that 
\begin{align}
F_\rho \sim A(M_\alpha)\, e^{-a \rho}\sim m_{3/2}\,,
\end{align}
as, for example, in KKLT. For the case of a condensing $SU(N_\text{c})$ gauge theory with a single meson $M$, one finds ${A(M)= (N_\text{c}-1)M^{-2/(N_\text{c}-1)}}$. The meson is then stabilized by the interplay of this instanton term and its mass term, as explained in detail in \cite{Dudas:2005vv}. Evidently, the instanton term is responsible for a large vev of $M$, which can be expressed as
\begin{align}\label{eq:wequation}
M \sim \frac{F_\rho}{m_M}\,.
\end{align}
Given a meson mass $m_M \sim m_{3/2}$, the minimum lies at $M \sim 1$ in Planck units. Therefore, if the constraint~\eqref{eq:fdterms} holds, the D-term contribution of the meson exceeds the size of the field-dependent FI term. This is inconsistent with having an effective realization of standard DHI.\footnote{While it may be possible to obtain an approximate version of DHI with an FI term generated by stabilized mesons, the analysis of such schemes is beyond the scope of this work.}

In order to find a possible way out of this apparent predicament, one may invoke schemes of modulus stabilization with $m_\rho \gg m_{3/2}$. An example of this kind may be given by the modulus stabilization via additional K\"ahler potential terms as proposed, for example, in \cite{ArkaniHamed:1998nu}. But even this is not a full solution to the problem, as stabilization of the mesons still requires a very large gravitino mass and DHI is spoiled by the large displacement of the meson, cf. Eq.~\eqref{eq:wequation}.

To summarize, in models where the field-dependent FI term drives inflation there is always an intimate connection between modulus stabilization and inflation. Generically, the modulus does not decouple from the dynamics of inflation. When trying to obtain the simple controllable scheme of DHI as an effective theory, a series of problems arises. These problems are related to the fact that inflation back-reacts on the modulus stabilization and vice versa. While we have shown that there is no straight-forward realization of DHI with a field-dependent FI-term, we can not exclude that it arises to some approximation by a very delicate engineering of the K\"ahler potential and the mechanism of modulus stabilization. 

In the following section we explore the possibility that inflation proceeds in the waterfall regime of hybrid inflation. In this case the FI term is canceled during inflation and modulus stabilization can be achieved via the breaking of $U(1)_\text{A}$. As discussed below, the obstacles mentioned above are absent in this case.

%

\section{F-Term Inflation from Field-Dependent FI Terms}
\label{sec:FtermFI}

In this section we present a string-inspired toy model of inflation with a field-dependent FI term which circumvents the problems discussed in the previous section. Inflation is driven by the F-term of one of the waterfall fields after the waterfall phase transition, as suggested in \cite{Buchmuller:2014rfa} and reviewed in Section~\ref{sec:ChaoticInf}. The resulting effective theory after integrating out the modulus and the heavy vector supermultiplet is identical to chaotic inflation with a quadratic potential. We first demonstrate our scheme of modulus stabilization and afterward discuss its coupling to inflation.

%
\subsection{Modulus Stabilization}
\label{sec:ModelModStab}

We consider a setup in which the modulus $\rho$ has the gauge-invariant superpotential
\begin{align}\label{eq:WModStabModel}
 W = \chi_+ \left(\m^2\, e^{- \rho /\delta_{\text{GS}}} -m \m \right) \,,
\end{align}
where $\chi_+$ and $\phi_-$ are chiral superfields with $U(1)_\text{A}$ charge $+1$ and $-1$, respectively. Yukawa couplings suppressed by an instanton action like the one in Eq.~\eqref{eq:WModStabModel} arise, for example, in intersecting D-brane models, cf.~\cite{Blumenhagen:2009qh}. Modulus stabilization via Yukawa-type interactions has previously been studied in \cite{Brummer:2010fr}. We assume the no-scale K\"ahler potential
\begin{align}\label{eq:noscalematter}
 K = -3\log\left[\rho + \bar{\rho} -2\delta_\text{GS} \mathcal V -  \mb e^{-2 \mathcal V} \m - \bar \chi_+ e^{2 \mathcal V} \chi_+ \right]\,,
\end{align}
where, as before, $\mathcal V$ denotes the $U(1)_\text{A}$ vector supermultiplet. We remark that the choice of $K$ in Eq.~\eqref{eq:noscalematter} is convenient and well-motivated from a string theory perspective, but the specific form of $K$ does not affect the following discussions.

At the level of global supersymmetry it is clear that minimizing the F-term potential of $\chi_+$ stabilizes $\rho$ at a non-zero vev. To see things more clearly, let us perform the following field redefinitions,
\begin{align}
 \mathcal V &\rightarrow \mathcal V + \log \left( \frac{|\m|}{\sqrt{\delta_\text{GS}}}\right)\,,\\
 \rho &\rightarrow \rho + \delta_\text{GS}\log\left( \frac{\m}{\sqrt{\delta_\text{GS}}}\right)\,,\\
 \chi_+ &\rightarrow \frac{\sqrt{\delta_\text{GS}}}{\m} \,\chi_+\,.
\end{align}
With these redefinitions, $\phi_-$ is eliminated from the spectrum and we obtain, analogous to Section~\ref{sec:ChaoticInf}, the superfield version of unitary gauge. The K\"ahler potential in this frame becomes
\begin{align}
 K = -3\log\left[\rho + \bar{\rho} -2 \delta_\text{GS} \mathcal V -\left(  \delta_\text{GS} e^{-2 \mathcal V}  +\bar \chi_+ e^{2 \mathcal V} \chi_+ \right)\right]\,.
\end{align}
We can again integrate out $\mathcal V$ supersymmetrically by solving its equation of motion and obtain
\begin{align}
 \mathcal V = - \frac{|\chi_+|^2}{2\delta_\text{GS}} + \mathcal{O}\left( |\chi_+|^4 \right)\,,
\end{align}
at leading order in $\chi_+$. Using this solution we find for the effective super- and K\"ahler potential,
\begin{align}\label{eq:ModelW}
W &= \delta_\text{GS} \, \chi_+ \left( e^{- \rho /\delta_{\text{GS}}} - \frac{m}{\sqrt{\delta_\text{GS}}} \right) \,, \\ \label{eq:ModelK}
 K &= -3\log\left[\rho + \bar{\rho} - \delta_\text{GS} -|\chi_+|^2 + \frac{|\chi_+|^4}{2\delta_\text{GS}}  \right]\,.
\end{align}
The effective Lagrangian defined by Eqs.~\eqref{eq:ModelW} and \eqref{eq:ModelK} has a supersymmetric minimum at 
\begin{align}\label{eq:rhoMin}
\rho_0 \equiv -\delta_\text{GS} \log\left(\frac{m}{\sqrt{\delta_\text{GS}}}\right)\,, 
\end{align}
and $\chi_+=0$. In the vacuum, the mass of the canonically normalized modulus is given by
\begin{align}\label{eq:rhomass}
 m_\rho=\frac{m}{3\sqrt{\delta_\text{GS}}}=\frac{1}{3}\,e^{-\frac{3}{2\xi_\text{GS}}}\,,
\end{align}
which coincides with the mass of $\chi_+$. The fermionic components of $\rho$ and $\chi_+$ combine into a Dirac fermion. Notice that there is an intricate relation between the size of the effective FI term defined in Eq.~\eqref{eq:xiGS} and the mass of the modulus field. A large supersymmetric mass for the modulus can be achieved by an effective FI term close to the Planck scale, as it naturally appears in string compactifications. As an example, $m_{\rho}\sim M_\text{GUT}$ corresponds to $\xi_\text{GS} \sim 0.4$. In this parameter regime it is possible to stabilize $\rho$ at a large vev, $\rho_0 \gg 1$ in Planck units, to guarantee the validity of the supergravity approximation. We remark that, in this discussion, we treat $\delta_\text{GS}$ as a free parameter since additional chiral states may contribute to the anomaly without affecting our discussion.

%
\subsection{Chaotic Inflation}

Consider the system defined by the superpotential 
\begin{align}\label{eq:ModelCombW}
 W = \chi_+ \left(\m^2 e^{- \rho / \delta_{\text{GS}}} - m \m \right) + \lambda \I \p \m\,,
\end{align}
which is obtained by adding the superpotential of DHI to the modulus sector discussed in Section~\ref{sec:ModelModStab}, identifying the waterfall field $\phi_-$ with the field which renders the instanton term gauge-invariant. As in Section~\ref{sec:Inflation}, $\varphi$ is protected by a shift symmetry in the K\"ahler potential and its real part is the inflaton field. The K\"ahler potential reads
\begin{align}\label{eq:ModelCombK}
 K = -3\log\left[\rho + \bar{\rho} - |\m|^2 - |\p|^2 - |\chi_{_+}|^2 + (\I- \Ib)^2 \right]\,.
\end{align}
Here we implicitly assume that the inflaton is part of the matter sector of a possible string theory embedding. As an example, it could be associated with a Wilson line scalar with the shift symmetry being a consequence of higher-dimensional gauge invariance. A discussion of the generic obstacles of large-field inflation in string theory by means of our toy model is, however, beyond the scope of this paper. For a recent discussion of large-field inflation with Wilson lines, see \cite{Marchesano:2014mla}. 
 
Along the lines of Section~\ref{sec:ChaoticInf} we can absorb $\phi_-$ into the vector superfield, which we integrate out supersymmetrically to obtain the effective theory for the remaining degrees of freedom. We find the following effective superpotential and K\"ahler potential,
\begin{align}
 W &= \delta_\text{GS} \,\chi_+ \left( e^{- \rho /\delta_{\text{GS}}} - 3 m_{\rho} \right)  + \lambda \sqrt{\delta_\text{GS}}\, \I  \p\,, \\
 K &= -3\log\left[\rho + \bar{\rho} - \delta_\text{GS} - |\p|^2 - |\chi_{_+}|^2 +\frac{(|\p|^2+|\chi_{_+}|^2)^2}{2\delta_\text{GS}}+ (\I- \Ib)^2 \right]\,,
\end{align}
where we have used Eq.~\eqref{eq:rhomass} to express $m$ in terms of the modulus mass. In case $\rho$ is stabilized at a scale far above the Hubble scale during inflation, it decouples from the dynamics and can be integrated out together with $\chi_+$, see Section~\ref{sec:ModelModStab}. The resulting effective theory for $\varphi$ and $\p$ can then be described by
\begin{align}\label{eq:ModelEff}
W= \hat{m} \hat{\I} \hat{\phi}_{_+} \,, \qquad K =  |\hat{\phi}_{_+}|^2- \frac{|\hat{\phi}_{_+}|^4}{2\xi_{\text{GS}}} - \frac{(\hat{\I}- \bar{\hat{\I}})^2}{2}\,,
\end{align}
where we have introduced the mass parameter $\hat{m}= \lambda \sqrt{\xi_\text{GS}}/3 \sqrt6$ and the canonically normalized superfields $\hat{\phi}_{_+}=\phi_{_+}\sqrt{3/(2\rho_0 - \delta_\text{GS})}$ and $\hat{\I}=\I\,\sqrt{6/(2\rho_0 - \delta_\text{GS})}$. Evidently, by integrating out all heavy degrees of freedom we have obtained the standard realization of chaotic inflation as an effective theory. 

However, for finite $m_\rho$ a small correction to the predictions of chaotic inflation arises due to a displacement of the modulus during inflation. Integrating out $\rho$ at its shifted vev induces an inflaton-dependent correction to the scalar potential. The leading-order correction can be found by expanding $V$ around $\rho_0$, i.e.,
\begin{align}\label{eq:ModCorrPot}
V = \frac{1}{(\rho + \bar \rho - \delta_\text{GS})^2} \Big[ \hat{m}^2\hat{\I}^2(2 \rho_0-\delta_\text{GS})^2+3 m_{\rho}^2 \,| \rho -\rho_0|^2 \Big ]\,,
\end{align}
where we have set $\hat \varphi = \bar{\hat \varphi}$ along the inflationary trajectory. Minimizing this expression with respect to $\rho$ gives
\begin{align}
 \rho -\rho_0 =  \frac{4\rho_0-2\delta_\text{GS}}{3m_{\rho}^2}\;\hat{m}^2\hat{\I}^2 \,.
\end{align}
By plugging this result back into Eq.~\eqref{eq:ModCorrPot} we can obtain an effective potential for the inflaton field. The correction can be conveniently expressed as a power series in $H/m_\rho$ along the lines of \cite{Buchmuller:2014vda}. Including the leading-order term we obtain
\begin{align}\label{eq:InfEffPot}
V = V_0 \left(1 - \frac{4}{3}\,\frac{V_0}{m_{\rho}^2}\right)\,,
\end{align}
with $V_0 = \hat m^2 \hat \varphi^2$. Notice that the numerical coefficient of the correction term differs from the result obtained in \cite{Buchmuller:2014vda} due to the different choice of K\"ahler potential. As naively expected, the correction induced by the shift of the modulus completely disappears in the limit where $\rho$ is infinitely heavy. For $m_\rho > H$ the second term in Eq.~\eqref{eq:InfEffPot} may have a controllable effect on the inflationary observables. The investigation of this effect is, however, beyond the scope of this note.

Notice that modulus stabilization in the effective theory implies constraints on the initial conditions of the system. In particular, inflation can not begin at arbitrarily large field values of $\hat \varphi$. To ensure that $\rho$ remains stabilized in the entire cosmological history, the energy density of the universe must never exceed the modulus mass. This is a conceptual subtlety which remains to be addressed in many effective theories of inflation with moduli stabilization.

%

\section{Conclusion}
\label{sec:Conclusion}

In this note we have pointed out an intimate connection between D-term hybrid inflation and chaotic inflation, related to Fayet-Iliopoulos terms of different origins. DHI driven by a constant FI term proceeds in the regime where a $U(1)$ gauge symmetry is unbroken, whereas chaotic inflation can be the effective theory after the $U(1)$ phase transition. In this picture, the mass of the inflaton is generated by a Yukawa interaction. The stabilizer field required in a supergravity realization of chaotic inflation can be identified with one of the waterfall fields. The non-minimal K\"ahler potential, which is usually introduced by hand to decouple the stabilizer from inflation, has its origin in a $U(1)$ gauge interaction. We explicitly show that it is obtained by integrating out the heavy vector superfield of the broken $U(1)$.

From a UV perspective constant FI terms pose a potential problem. Therefore, we have investigated the implementation of DHI and the aforementioned chaotic inflation setup with a field-dependent FI term. In the presence of an anomalous $U(1)$ symmetry, the latter is related to a string modulus which contains the Green-Schwarz axion. We studied the question whether a field-dependent FI term can play the role of its constant counterpart in the early universe. We found that in settings where the field-dependent FI term provides the vacuum energy of inflation, there is always a strong interplay between modulus stabilization and inflation. In models where the charged modulus appears in the superpotential, properly accounting for gauge invariance prevents a decoupling of the modulus from the dynamics of inflation. While successful D-term inflation may be obtained, the resulting scheme does not resemble the simple controllable pattern of DHI.

This leads us to consider inflation in the broken phase of the $U(1)$ symmetry. We have shown that the $U(1)$ phase transition, triggered by the field-dependent FI term, can lead to a successful realization of chaotic inflation -- analogous to the case of a constant FI term. In this setting the modulus couples to the $U(1)$ breaking field through an instanton-suppressed Yukawa coupling. After the breaking of the $U(1)$ symmetry a supersymmetric modulus mass term arises whose size is controlled by the field-dependent FI term. With the FI term close to the Planck scale, as expected in realistic string constructions, $m_\rho$ exceeds the Hubble scale of inflation. In this case the back-reaction of the modulus on the inflaton potential is under control.

Eventually, a picture emerges in which string moduli do not participate in supersymmetry breaking. This appears very attractive from a phenomenological perspective as it allows for low-energy supersymmetry breaking without light moduli.  

%

\subsection*{Acknowledgments}
We are grateful to Emilian Dudas, Kwang Sik Jeong, Robert Richter, Fabian R\"uhle, and Alexander Westphal for stimulating discussions, and to Wilfried Buchm\"uller for helpful comments on the manuscript. This work has been supported by the German Science Foundation (DFG) within the Collaborative Research Center 676 ``Particles, Strings and the Early Universe''. The work of CW is supported by a scholarship of the Joachim Herz Stiftung.

\end{document}